\documentclass[a4paper,reqno,11pt]{amsart}
\begin{document}
\title[A rotating three component perfect fluid] {A rotating three component perfect fluid source and its junction with empty space-time}
\author[R.J. Wiltshire] {R.J. Wiltshire \\ \\ \tiny{University of Glamorgan \\ Pontypridd CF37 1DL, \\ Wales, UK \\ \\ email: rjwltsh@glam.ac.uk}}
\maketitle
\begin{center}
	\Small{\textsc{\today}}
\end{center}

\begin{abstract}
The Kerr solution for empty space-time is presented in an ellipsoidally symmetric coordinate system and it is used to produce generalised ellipsoidal metrics appropriate for the generation of rotating interior solutions of Einstein's equations. It is shown that these solutions are the familiar static perfect fluid cases commonly derived in curvature coordinates but now endowed with rotation. These are also shown to be potential fluid sources for not only Kerr but also  Kerr-de Sitter empty space-time. The approach is further discussed in the context of T-solutions of Einstein's equations and the vacuum T-solution outside a rotating source is presented. The interior source for these solutions is shown not to be a perfect fluid but rather an anisotropic three component perfect fluid for which the energy momentum tensor is derived. The Schwarzschild interior solution is given as an example of the approach.
\end{abstract}
%INTRODUCTION
\section{Introduction}
The problem of finding a suitable rotating fluid source which joins smoothly to the asymptotically flat, vacuum gravitational field first published by Kerr \cite{kerr} remains of considerable interest in general relativity. Candidate closed form solutions are extremely rare and only for the important case of thin super-massive rotating discs, supported by internal pressure have analytic sources been derived as is the case of Pichon and Lynden-Bell \cite{pich}.  The interior solution of Wahquist \cite{wahl} solution has been the subject of considerable interest as a possible, although physically unrealistic solution. However recently Bradley \textit{et al} \cite{brad} demonstrated that a fit to the Kerr solution was not possible even though Sarnobat and Hoenselaers \cite{sarn} showed using a series approximation method,  that a fit was possible up to second order terms in an angular velocity parameter.  The series approach to the matching problem is more common in the literature. For example Hartle \cite{hart} uses a second order perturbation technique to describe the slow rotation of equilibrium configurations of cold stars with constant angular velocity whilst Wiltshire \cite{wilt2} generalises this approach using Darmois \cite{darm} boundary conditions (see also Bonnor and Vickers \cite{bonn}). 

In an alternative approach it has been suggested that imperfect fluid solutions of Einstein's equations may produce suitable sources for Kerr empty space-time. Both Hernandez \cite{hern} and Lettelier \cite{lett} have produced possible alternative energy momentum tensors although the latter is more realistic, offering as it does, a clear interpretation of anisotropy in fluid sources based upon an invariant combination of two component perfect fluids. These tensors have not been used significantly in the context of rotation. However non-rotating anisotropic solutions of Einstein's equations based upon the energy momentum tensor of Lettelier have been derived by many authors. For example, Mak and Harko \cite{mak} and Sharma and Maharaj \cite{sha} have recently derived solutions with a view to determining physically realistic fluid sources matching the exterior Schwarzschild solution. Since the pioneering work on anisotropy by Bowers and and Liang \cite{bow} there has been considerable study of relativistic anisotropy and reviews have been presented by Delgaty and Lake \cite{del} and Herrera and Santos \cite{her}. The notion of anisotropy in rotating sources will be shown to be of considerable significance in the discussion below.

Although mathematical descriptions  of rotation frequently centre upon axially symmetric stationary metrics, for example Stephani \cite{step} and Islam \cite{isl} it has been shown by Krasi\'{n}ski \cite{kras} that the axially symmetric ellipsoidal form may hold advantages in terms the description of relativistic fluid interiors, a fact reinforced through its close resemblance to the geometry associated with rotating Newtonian fluids. More recently both  R\'{a}cz \cite{racz} and Zsigrai \cite{zsi}  have generalised the ideas of Krasinski and provided a rigorous mathematical basis for their study although they were not exploited in terms of generating rotating fluid interiors.

It is the aim here to build upon this work and demonstrate how the ellipsoidal form may be used to describe rotating fluid interiors in such a way that they form a natural match to both Kerr analso Kerr-de Sitter vacuum space-times. This is achieved by observing the close links between ellipsoidal metrics and the normal curvature forms used to determine static non-rotating solutions of Einstein's equations. The non-rotating static solutions are sometimes referred to as  R-solutions Novikov,  \cite{nov} Novikov and Zeldovich \cite{zeld} and McVittie and Wiltshire \cite{wilt1} and these have 'mirrror' solutions, valid within the Schwarzschild radius, known as T-solutions which will also be shown to have a rotating ellipsoidal counterpart. In this way the T-vacuum counterpart to the Kerr and the Kerr-de Sitter solution will be presented in Section \ref{tsolution}.  An appropriate rotating fluid source suitable for matching ellipsoidally symmetric geometry is discussed in Sections \ref{source} and \ref{empty}. In particular it will be shown that suitable model sources may be constructed using an extension of the Lettelier anisotropic approach to a tensor constructed from three component fluid sources. Details of the matching process are presented and applied to the Schwarzschild Interior solution endowed with rotation.

%ELLIPSOIDAL AND CURVATURE COORDINATE SYSTEMS
\section{ Ellipsoidal and curvature coordinate systems}\label{ein}

In the following a metric with axially ellipsoidal symmetry will be discussed in the form  
\begin{eqnarray}
ds^2 &=& \nu(dt+f d\phi)^2  \nonumber \\
	 && -\frac{1}{\nu}(\lambda d\rho^2 +(\rho^2+a^2\cos^2\theta)d\theta^2 +(\rho^2+a^2)\sin^2\theta d\phi^2) ,
	 \label{metric}
\end{eqnarray}
where $f=f(\rho,\theta), \lambda=\lambda(\rho,\theta), \nu=\nu(\rho,\theta)$. It will be used to describe the gravitational field of a rotating anisotropic fluid source.  Axially symmetric ellipsoidal geometry has been described by for example Krasi\'{n}ski \cite{kras} and Zsigrai \cite{zsi} and an important property in the following is that the  Kerr vacuum solution of Einstein's equations is given by: 
\begin{equation}\label{fkerr}
	f=f_{Kerr}={{2\sin^2\theta am(\sqrt{m^2+\rho^2}+m)}\over{a^2\cos^2\theta+\rho^2}},
\end{equation}
\begin{equation}\label{lkerr}
	\lambda=\lambda_{Kerr}={{\rho^2(a^2\cos^2\theta+\rho^2)}\over{(a^2+\rho^2)(m^2+\rho^2)}},
\end{equation}
\begin{equation}\label{nkerr}
	\nu=\nu_{Kerr}={{a^2\cos^2\theta+\rho^2}\over{2m(\sqrt{m^2+\rho^2}+m)+a^2\cos^2\theta+\rho^2}},
\end{equation}
where he constant $a$ is the angular velocity parameter. Equations (\ref{fkerr}) to (\ref{nkerr}) may be cast into the more familiar Boyer-Lindquist \cite{boy} form by setting $\rho^2=r^2-2mr$ as has been shown by R\'{a}cz \cite{racz}.

To determine possible interior solutions of Einstein's equations consider the generalised form of equations (\ref{fkerr}) to (\ref{nkerr}) but with the similar structure:

\begin{equation}\label{fsol}
	f={{aZ \sin^2\theta }\over{a^2\cos^2\theta+\rho^2}},
\end{equation}
\begin{equation}\label{lsol}
	\lambda=\frac{Y(a^2\cos^2\theta+\rho^2)}{\rho^2+a^2},
\end{equation}
\begin{equation}\label{nsol}
	\nu={{a^2\cos^2\theta+\rho^2}\over{X+a^2\cos^2\theta}},
\end{equation}
where $X=X(\rho)$, $Y=Y(\rho)$ and $Z=Z(\rho)$ are arbitrary functions of $\rho$. 

In general the Einstein tensor $G^{a}_{b}$ for this metric has non-zero components $G^{1}_{1} ,G^{1}_{2},G^{2}_{1}, G^{2}_{2}, G^{3}_{3}, G^{3}_{4}, G^{4}_{3},G^{4}_{4}$. However explicit calculation shows that
\begin{equation}\label{g12}
	 G^{1}_{2}=0=G^{2}_{1},
 \end{equation}
 is satisfied whenever:
\begin{eqnarray}
		a^2\cos^2\theta \left(  ZZ_{\rho}-XX_{\rho}+\rho^2 X_{\rho}+2\rho X-2\rho^2 \right)   \nonumber \\
+\rho^2ZZ_{\rho}-2\rho Z^2-\rho^2XX_{\rho}+\rho^4X_{\rho}+2\rho X^2-2\rho^3X=0, \label{g12extra} 
\end{eqnarray}
where a suffix indicates a derivative. This has a particular solution
\begin{equation}\label{wilt}
	Z=X-\rho^2,
\end{equation}
which will be used in equation (\ref{fsol}).

When the angular velocity parameter $a=0$, the metric (\ref{metric}) may be transformed to the normal curvature coordinate system:
\begin{equation}\label{curve}
	ds^2=e^{2\alpha}dt^2-e^{2\beta}dr^2-r^2d\theta^2-r^2\sin^2\theta^2d\phi^2,
\end{equation}
where, $\alpha=\alpha(r)$ and $\beta=\beta(r)$, by means of the coordinate transformation:
\begin{equation}\label{standard}
	\rho=re^\alpha, \hspace{0.5in} X=r^2, \hspace{0.5in} Y=\frac{e^{2\beta}}{(1+r\alpha_{r})^2}.
\end{equation}
Using this transformation the metric (\ref{metric}) can be written in the following ellipsoidal-curvature form
\begin{eqnarray}
	ds^2 &=& \bar{\nu}(dt+\bar{f} d\phi)^2  \nonumber \\
		& -&  {{1}\over{\bar{\nu}}}(\bar{\lambda} dr^2+(r^2e^{2\alpha}+a^2\cos^2\theta) d\theta^2+(r^2e^{2\alpha}+a^2) \sin^2\theta d\phi^2), \nonumber \\
	 \label{ellip_curve}
\end{eqnarray}
where $\bar{f}=\bar{f}(r,\theta), \bar{\lambda}=\bar{\lambda}(r,\theta), \bar{\nu}=\bar{\nu}(r,\theta)$ and:

\begin{equation}\label{fbar}
	\bar{f}=\frac{ar^2(1-e^{2\alpha})\sin^2\theta}{r^2e^{2\alpha}+a^2\cos^2\theta},
\end{equation}
\begin{equation}\label{lbar}
	\bar{\lambda}=\frac{(r^2e^{2\alpha}+a^2\cos^2\theta)e^{2(\alpha+\beta)}}{r^2e^{2\alpha}+a^2},
\end{equation}
\begin{equation}\label{nbar}
	\bar{\nu}=\frac{r^2e^{2\alpha}+a^2\cos^2\theta}{r^2+a^2\cos^2\theta}.
\end{equation}
The Kerr solution described in Boyer-Lindquist coordinates can be obtained by substituting
\begin{equation}\label{vac}
e^{2\alpha}=e^{-2\beta}=1-\frac{2m}{r},
\end{equation}
of the Schwarzschild solution into (\ref{ellip_curve}).  

Notice that in the case when the cosmological constant $\Lambda \ne 0$, the Kerr vacuum space-time generalises to the Kerr-de Sitter solution first found by Carter\cite{cart}. For such cases the metric (\ref{ellip_curve}), with equations (\ref{fbar}) to (\ref{nbar}), may be generalised in a way which retains the property that $G^1_2=0=G^2_1$ and reduces to (\ref{curve}) when the angular velocity parameter $a=0$. The result can be written compactly in the form:
\begin{eqnarray}
ds^2 &=& \frac{\Delta_r(dt-a\sin^2\theta d\phi)^2}{\kappa^2 h^2}-\frac{\Delta_{\theta}\sin^2\theta(adt-(r^2+a^2)d\phi)^2}      {\kappa^2 h^2} \nonumber \\
& & -h^2\left(\frac{e^{2(\alpha+\beta)}dr^2}{\Delta_r}+\frac{d\theta^2}{\Delta_\theta}\right), \label{desitter}
\end{eqnarray}
where,
\begin{eqnarray}
\Delta_r &=& a^2\left(1-\frac{\Lambda r^2}{3} \right) +r^2 e^{2\alpha}, \hspace{0.5in}
\Delta_\theta = 1+\frac{\Lambda a^2 \cos^2\theta}{3}, \nonumber \\
h^2 &=& r^2+a^2\cos^2\theta, \hspace{1.25in} \kappa = 1+ \frac{\Lambda a^2}{3}. \label{desitter1}
\end{eqnarray}
The metric in the form (\ref{desitter}) is a generalisation of Kerr-de Sitter vacuum solution presented by Akcay and Matzner\cite{akcay}. This particular form can be found by setting:
\begin{equation}\label{vac1}
e^{2\alpha}=e^{-2\beta}=1-\frac{2m}{r}-\frac{\Lambda r^2}{3},
\end{equation}
in (\ref{desitter}) and (\ref{desitter1}).  
  
Consider now potential fluid sources with energy momentum tensor $T^a_{b}$ for the metric (\ref{ellip_curve})  Einstein's equations in the form:
\begin{equation}\label{eineqs}
	G^{a}_{b}=8\pi T^{a}_{b}.
\end{equation}

In principle any solution of Einstein's equations corresponding to the non-rotating curvature system (\ref{curve}) can be substituted into the ellipsoidal-curvature form (\ref{ellip_curve}), (\ref{fbar}) to (\ref{nbar})  and it will satisfy the condition $G^{1}_{2}=0=G^{2}_{1}$. However whilst metrics of the form (\ref{curve}) are also often used (for example Stephani \textit{et al} \cite{step}) in the description of perfect fluid sources
\begin{equation}\label{fluid1}
      T^{a}_{b}=(\varrho+p)U^aU_b-\delta^{a}_{b}p, \hspace{0.5in}U^{a}U_{a}=1,
\end{equation}
these are not also compatible with a source described by a rotating ellipsoidal-curvature solution given by (\ref{ellip_curve}). For example isotropy of pressure $p$ is not possible in these cases. 

Similar  comments are also valid for a generalised source consisting of a two component perfect fluid presented by Lettelier \cite{lett} and described by
\begin{equation}\label{fluid2}
      T^{a}_{b}=(\varrho+p)U^aU_b+(\sigma-p)\chi^a\chi_b -\delta^{a}_{b}p,
\end{equation}
where,
\begin{equation}\label{fluid2a}
	U^{a}U_{a}=1, \hspace{0.5in} \chi^{a}\chi_{a}=-1, \hspace{0.5in} U^{a}\chi_{a}=0.
\end{equation}

Thus recent examples, Mak and Harko \cite{mak}, Sharma and Maharaj \cite{sha} of anisotropic two component fluid sources (\ref{fluid2}) described by the metric (\ref{curve}) can also be substituted into (\ref{ellip_curve}) to correspond to a rotating source with $G^{1}_{2}=0=G^{2}_{1}$.  However the source would no longer be described by equation (\ref{fluid2}).

An appropriate anisotropic fluid source for the metric (\ref{ellip_curve}) will be discussed in Section \ref{source}.

% ELLIPSOIDAL FORMS OF THE T-SOLUTION
\section{Ellipsoidal forms of the T-solution}\label{tsolution}

However before this discussion note that the metric (\ref{curve}) also has a T-solution counterpart as has been described by Novikov \cite{nov}, McVittie and Wiltshire \cite{wilt1} of the form:
\begin{equation}\label{curve_T}
ds^2=e^{2V}dt^2-e^{2W}dr^2-t^2d\theta^2-t^2\sin^2\theta^2d\phi^2,
\end{equation}
where $V=V(t)$ and $W=W(t)$ and the metrics (\ref{curve}) and (\ref{curve_T}) are related by means of the transformation:
\begin{equation}\label{trans_T}
r=\bar{t}, \hspace{0.5in} t=\bar{r}, \hspace{0.5in} e^{2\alpha(r)}=-e^{2W(\bar{t})}, \hspace{0.5in} e^{2\beta(r)}=-e^{2V(\bar{t})},
\end{equation}
so that for empty space-time $1-2m/r=-(2m/\bar{t}-1)$. The bars are omitted in the notation to give (\ref{curve_T}).

	It is interesting to use the approach developed in the previous section to develop metrics applicable to rotating interior T-solutions and further to determine a T-solution form of Kerr empty space-time. Hence using the approach of the previous section together with the transformation (\ref{trans_T}) the ellipsoidal-curvature metric (\ref{ellip_curve}) may be adapted to a form appropriate for use in a T-region as follows:
\begin{eqnarray}
	ds^2 &=& \frac{\lambda^{T}}{\nu^{T}}dt^2-\nu^{T}(dr+f^{T} d\phi)^2  \nonumber \\
		& -&  {{1}\over{\nu^{T}}}((t^2e^{2W}-a^2\cos^2\theta) d\theta^2+(t^2e^{2W}-a^2) \sin^2\theta d\phi^2), \nonumber \\
	 \label{ellip_T}
\end{eqnarray}
where now $f^{T}=f^{T}(t,\theta), \lambda^{T}=\lambda^{T}(t,\theta), \nu^{T}=\nu^{T}(t,\theta)$ and:
\begin{equation}\label{fts}
	f^{T}=\frac{at^2(1+e^{2W})\sin^2\theta}{t^2e^{2W}-a^2\cos^2\theta},
\end{equation}
\begin{equation}\label{lts}
	\lambda^{T}=\frac{(t^2e^{2W}-a^2\cos^2\theta)e^{2(V+W)}}{t^2e^{2W}-a^2},
\end{equation}
\begin{equation}\label{nts}
	\nu^{T}=\frac{t^2e^{2W}-a^2\cos^2\theta}{t^2+a^2\cos^2\theta}.
\end{equation}
Note that in equations (\ref{ellip_T}) to (\ref{nts}) each of the components of the Einstein tensor compute to zero except for $G^1_1$, $G^2_2$, $G^3_3$, $G^4_4$, $G^1_3$ and $G^3_1$ which are non zero in general. 

For example the non-rotating T-solution due to McVittie and Wiltshire \cite{wilt1}:
\begin{equation}\label{tex}
e^{-2V}=\frac{4}{3} \left(\frac{2m}{t}-1 \right)^{-1}, \hspace{0.5in} e^{2W}=\frac{1}{t} ,\hspace{0.5in} p=\frac{\varrho}{3}=\frac{1}{t^2},
\end{equation}
may be substituted into the rotating system (\ref{ellip_T}), and (\ref{fts} to (\ref{nts}) with the result that  $G^{1}_{2}=0=G^{2}_{1}$. Potential sources for solutions of this type will be provided in the next section.

In the particular case of the vacuum T-solution in  (\ref{curve_T}): 
\begin{equation}\label{vac_T}
	e^{2W}=e^{-2V}=\frac{2m}{t}-1,
\end{equation}
direct substitution of (\ref{vac_T}) into equations (\ref{ellip_T}) to (\ref{nts}) demonstrates that:
\begin{equation}\label{Tkerr}
	G^a_{b}=0,
\end{equation}
for all values of $t$ and $\theta$ thus the resulting vacuum solution of Einstein's equations is T-region analogue of the Kerr solution. 

In the case when the cosmological constant $\Lambda \ne 0$, the metric (\ref{ellip_T}), with equations (\ref{fts}) to (\ref{nts}), may be generalised in a way which retains the property that $G^1_2=0=G^2_1$. and reduces to (\ref{curve_T}) when $a=0$. The result is:
\begin{eqnarray}
ds^2 &=& \frac{\Delta^T_t(dr-a\sin^2\theta d\phi)^2}{\kappa^2 (h^{T})^2}-\frac{\Delta^T_{\theta}\sin^2\theta(adr-(t^2+a^2)d\phi)^2}      {\kappa^2 (h^{T})^2} \nonumber \\
& & -(h^{T})^2\left(\frac{e^{2(V+W)}dt^2}{\Delta^T_t}+\frac{d\theta^2}{\Delta^T_\theta}\right), \label{tdesitter}
\end{eqnarray}
where,
\begin{eqnarray}
\Delta^T_t &=& a^2\left(1-\frac{\Lambda t^2}{3} \right) -t^2 e^{2W}, \hspace{0.5in}
\Delta^T_\theta = 1+\frac{\Lambda a^2 \cos^2\theta}{3}, \nonumber \\
(h^{T})^2 &=& t^2+a^2\cos^2\theta, \hspace{1.25in} \kappa = 1+ \frac{\Lambda a^2}{3}. \label{tdesitter1}
\end{eqnarray}
Finally, the corresponding Kerr-de Sitter vacuum T-solution can be found by substituting:
\begin{equation}\label{tvac1}
e^{2W}=e^{-2V}=\frac{\Lambda t^2}{3}+\frac{2m}{t}-1,
\end{equation}
in (\ref{desitter}) and (\ref{desitter1}). The T-solutions will not be investigated further here.

%THREE COMPONENT PERFECT FLUID TENSOR
\section{A three component perfect fluid tensor}\label{source}

In this section suitable fluid sources will be provided for the rotating systems described by (\ref{ellip_curve}), (\ref{desitter}) and (\ref{ellip_T}), (\ref{tdesitter}) by generalising the energy momentum tensor (\ref{fluid2}) from a two component to a three component fluid tensor. This can be achieved by considering the following three perfect fluids which will form the components of the new tensor:
\begin{equation}\label{tn10}
	T_{A}^{ab}(u)=Au^au^b-g^{ab}p_A,
\end{equation}
\begin{equation}\label{tn20}
	T_{B}^{ab}(v)=Bv^av^b-g^{ab}p_B,
\end{equation}
\begin{equation}\label{tn30}
	T_C^{ab}(w)=Cw^aw^b-g^{ab}p_C,
\end{equation}
where 
\begin{equation}\label{tn32c}
       A=\rho_{A}+p_{A} ,\hspace{0.5in} B=\rho_{B}
       +p_{B}, \hspace{0.5in} C=\rho_{C}+p_{C},
\end{equation}
and where $\rho_A,\rho_B, \rho_C$ are the respective internal energy densities and $p_A, p_B,p_C $ are the respective internal pressures. The fluid velocity four vector are $u^a, v^a$ and $w^a$ and satisfy:
\begin{equation}\label{tn35}
         u^au_a=1, \hspace{0.7in} v^av_a=1,
          \hspace{0.7in} w^aw_a=1.
\end{equation}

In an extension of the method introduced by Letelier \cite{lett} the composite three component tensor $T^{ab}(u,v,w)$ is now formed such that:
\begin{equation}\label{tn40}  
      T^{ab}(u,v,w)=T_A^{ab}(u)+T_B^{ab}(v)+T_C^{ab}(w),
\end{equation}
and which may also be expressed in the form
\begin{equation}\label{eq1}
      T^{ab}=\varrho U^a U^b+S^{ab}, \hspace{0.5in} S^{ab}U_{b}=0,
\end{equation}
where $\varrho$ expresses the energy density of the composite tensor and $S^{ab}$ the stress tensor component

To achieve the form (\ref{eq1}) note that equation (\ref{tn40}) with equations (\ref{tn10}) to (\ref{tn30}) is a quadratic form and is invariant under the transformation:
\begin{equation}\label{tn42}
    \left[
	    \begin{array}{c}
	          u^a \\ v^a \\ w^a 
	     \end{array}
    \right]
     =R_3(\hat{\psi})R_2(\hat{\theta})R_1(\hat{\phi})
    \left[
	   \begin{array}{c}
	        \bar{u}^a \\ \bar{v}^a \\ \bar{w}^a 
	 \end{array}
    \right],
\end{equation}
where $R_1, R_2, R_3$ are the Euler rotation matrices:
\begin{equation}\label{tn45}
	R_1(\hat{\phi})=
	\left[
		\begin{array}{ccc}
			 1 & 0 &   0\\
			 0 & \cos\hat{\phi}  &  -\sqrt{{C}\over{B}}\sin\hat{\phi} \\
			 0 & \sqrt{{B}\over{C}}\sin\hat{\phi}  &  \cos\hat{\phi} 
		\end{array}
	\right],
\end{equation}
\begin{equation}\label{tn45a}
R_2(\hat{\theta})=
\left[
\begin{array}{ccc}
 \sqrt{{A}\over{C}}\cos\hat{\theta} & 0 &   \sin\hat{\theta} \\
 0 & 1  &  0 \\
 -\sqrt{{A}\over{C}}\sin\hat{\theta} & 0  &  \cos\hat{\theta} 
\end{array}
\right],
\end{equation}
\begin{equation}\label{tn45b}
R_3(\hat{\psi})=
\left[
\begin{array}{ccc}
 \sqrt{{C}\over{A}}\cos\hat{\psi} & -\sqrt{{B}\over{A}}\sin\hat{\psi} &   0 \\
 \sqrt{{C}\over{B}}\sin\hat{\psi} & \cos\hat{\psi}  &  0 \\
 0 & 0  &  1
\end{array}
\right].
\end{equation}
Thus from equation (\ref{tn40}) with equations (\ref{tn10}) to (\ref{tn30}) and the transformation (\ref{tn42}) it follows that:
\begin{eqnarray}
      T^{ab}(u,v,w) &=& T^{ab}(\bar{u},\bar{v},\bar{w}) 
                                                            \nonumber  \\
                         &=& A\bar{u}^a\bar{u}^b+B\bar{v}^a\bar{v}^b
                                +C\bar{w}^a\bar{w}^b-g^{ab}p ,               
                                                             \label{tn50}
\end{eqnarray}
where,
\begin{equation}\label{tn60}
       p=p_A+p_B+p_C.
\end{equation}
Note  that equation (\ref{tn50}) together with (\ref{tn35}) implies that:
\begin{equation}\label{aa}
       \bar{u}^a\bar{u}_a=-{{C}\over{A}}\bar{w}^a\bar{w}_a
	+{{C}\over{A}}
       -{{B}\over{A}}\bar{v}^a\bar{v}_a+{{B}\over{A}}+1.
\end{equation}
Thus when $\bar{v}^a$ and $\bar{w}^a$ are transformed to be spacelike so that $\bar{v}^a\bar{v}_a<0$ and $\bar{w}^a\bar{w}_a<0$ then it follows from (\ref{aa}) that $\bar{u}^a$ is timelike so that $\bar{u}^a\bar{u}_a>0$. Such a transformation is possible by requiring orthogonality of $\bar{u}^a, \bar{v}^a, \bar{w}^a$ so that
\begin{equation}\label{bb}
	\bar{u}^a\bar{v}_a=0, \hspace{0.5in}
 	\bar{v}^a\bar{w}_a=0, \hspace{0.5in} \bar{w}^a\bar{u}_a=0.
\end{equation}
These conditions result in three equations for the Euler angles $\hat{\phi}, \hat{\theta}, \hat{\psi}$ which are lengthy and so are not presented here.

To present the three component fluid tensor (\ref{tn40}) in form of (\ref{eq1}) define:
\begin{equation}\label{tn70}
          U^a={\bar{u}^a\over{(\bar{u}^b\bar{u}_b)^{1/2}}}, 
     \hspace{0.5in}
         \chi^a={\tilde{v}^a\over{(-\tilde{v}^b\tilde{v}_b)^{1/2}}} ,
    \hspace{0.5in}
         \xi^a={\hat{w}^a\over{(-\hat{w}^b\hat{w}_b)^{1/2}}},
\end{equation}
where $U^a$ is a timelike vector and $\chi^a$, $\xi^a$ are spacelike vectors so that
\begin{equation}\label{tn75}
        U^aU_a=1, \hspace{0.5in}
        \chi^a\chi_a=-1, \hspace{0.5in}
        \xi^a\xi_a=-1,
\end{equation}
and which also satisfy the orthogonality conditions:
\begin{equation}\label{tn77}
       U^a\chi_a=0, \hspace{0.5in}
       U^a\xi_a=0, \hspace{0.5in}
       \chi^a\xi_a=0.
\end{equation}
With the further definition
\begin{equation}\label{tn80}
       \varrho=T^{ab}U_aU_b. \hspace{0.5in}
       \sigma=T^{ab}\chi_a\chi_b. \hspace{0.5in}
       \varpi=T^{ab}\xi_a\xi_b.
\end{equation}
it follows from (\ref{tn50}) and (\ref{tn70}) that
\begin{equation}\label{tn85}
        \varrho=A\bar{u}^k\bar{u}_k-p, \hspace{0.5in}
        \sigma=B\bar{v}^k\bar{v}_k+p, \hspace{0.5in}
        \varpi=C\bar{w}^k\bar{w}_k+p.
\end{equation}
Thus from  (\ref{tn80}) it can be seen that (\ref{tn50}) may be written in the form
\begin{equation}\label{tn100}
      T^{a}_{b}=(\varrho+p)U^aU_b+(\sigma-p)\chi^a\chi_b 
        +(\varpi-p)\xi^a\xi_b-\delta^{a}_{b}p.
\end{equation}
This of course is a natural extension to three perfect fluid components of the tensor first proposed by Lettelier  \cite{lett} for the two component perfect fluid tensor (\ref{fluid2}).

%The source for Kerr spacetime - THREE COMPONENT FLUID
\section{A particular three component fluid source for empty spacetime}\label{empty}
Consider now the conditions which are applicable such that the metric (\ref{ellip_curve}) for a rotating ellipsoidal system is compatible with the energy momentum tensor in the form (\ref{tn100}). First notice that the only non zero components of the Einstein tensor are $G^1_{1}, G^2_{2}, G^3_{3}, G^4_{4}$ and $G^3_{4}, G^4_{3}$ which means that the vectors $U^a, \chi^a, \xi^a$ defined in (\ref{tn77}) and (\ref{tn80}) must satisfy:
\begin{equation}\label{ueqn}
	U^{1}=0=U^{2}, \hspace{0.5in} U^3U_3+U^4U_4=1,
\end{equation}
\begin{equation}\label{ceqn}
		\chi^a=\sqrt{\frac{\nu}{\lambda}}\delta^a_{1}=\sqrt{ \frac{r^2e^{2\alpha}+a^2}{r^2+a^2\cos^2\theta}} e^{-(\alpha+\beta)}\delta^a_{1},
\end{equation}

\begin{equation}\label{xeqn}
	\xi^a=\sqrt{\frac{\nu}{r^2e^{2\alpha}+a^2\cos^2\theta}}\delta^a_{2}=\sqrt{\frac{1}{r^2+a^2\cos^2\theta}}\delta^a_{2}.\end{equation}

Thus from (\ref{tn100}) it can immediately be shown that:
\begin{equation}\label{pres1}
	\sigma=-T^{1}_{1}, \hspace{0.5in} \varpi=-T^2_2,
\end{equation}
and that the pressure $p$ satisfies the consistency relationship: 
\begin{equation}\label{cons}
	(T^3_3+p) (T^4_4+p)=T^3_4 T^4_3,
\end{equation}
It follows therefore that:
\begin{equation}\label{peqna}
	p=\frac{-T^3_3-T^4_4+(T^4_{4}-T^3_{3})\sqrt{(1+\epsilon^2_R}}{2},
\end{equation}
where
\begin{equation}\label{eps}
	\epsilon^2_{R}=\frac{4T^3_{4} T^4_{3}}{(T^4_{4}-T^3_{3})^2},
\end{equation}
and so evaluating $T^a_a$ in (\ref{tn100}) with (\ref{tn75}) it follows that:
\begin{equation}\label{densz}
\varrho=p+T^3_3+T^4_4 =\frac{T^3_3+T^4_4+(T^4_{4}-T^3_{3})\sqrt{(1+\epsilon^2_R}}{2}.
\end{equation}

In the particular case of slow rotation (\ref{peqna}) and (\ref{densz}) become:
\begin{eqnarray}
p &=& -T^3_{3}+\frac{T^3_{4}T^4_{3}}{T^4_{4}-T^3_{3}}+\cdots, \nonumber \\
\varrho &=& {}  T^4_{4}+\frac{T^3_{4}T^4_{3}}{T^4_{4}-T^3_{3}} +\cdots. \label{approxaa}  
\end{eqnarray}
Since both $T^3_{4}$ and $T^4_{3}$ are proportional to the angular velocity parameter $a$ the approximation (\ref{approxaa}) is valid upto and including terms in $a^2$.

Using the definition
\begin{equation}\label{dens}
	 \Omega = \frac{T^3_4}{p+\varrho}\hspace{0.5in} \Rightarrow\hspace{0.5in} U^3=\Omega U^4,
\end{equation}
it follows from the orthogonality condition  (\ref{ueqn}) that
\begin{equation}\label{omega}
	U^{4}=\frac{1}{\sqrt{g_{33}\Omega^2+2g_{34}\Omega+g_{44}}}. 
\end{equation}
Hence from the metric (\ref{ellip_curve}) or (\ref{ellip_T}) and Einstein's equations (\ref{eineqs}), the physical characteristics $U^a, \chi^a, \xi^a, \sigma, \varpi, p, \varrho$ of a three component perfect fluid may be computed from equations (\ref{ceqn}), (\ref{xeqn}), (\ref{pres1}), (\ref{peqna}), (\ref{densz}), (\ref{dens}) and (\ref{omega}). In principle the results are expressible in closed form even if practice it would rarely be attempted.

%Boundary Conditions

It is now convenient to apply Darmois junction conditions \cite{darm} which require the continuity of the first and second fundamental forms across a boundary surface $r=r_{b} $. 

In the case of the curvature coordinate system (\ref{curve}), without rotation, $a=0$, continuity of the first fundamental form implies: 
\begin{equation}\label{cond1}
\left \{  e^{2\alpha}\right \}_{r=r_{b}} =\left \{e^{-2\beta}\right \}_{r=r_{b}} =1-\frac{2m}{r_{b}},
\end{equation}
whilst continuity of the second fundamental form requires continuity of $\alpha_{r}$ across the $r=r_{b}$ which also implies a zero value for the pressure. In particular:
\begin{equation} \label{cond2}
	\left \{\frac{de^{2\alpha}}{dr} \right \}_{r=r_{b}} =\frac{2m}{r^2_{b}} \hspace{0.5in} \Rightarrow \hspace{0.5in}
	\left \{ \sigma(r_{b})=0 \right \}_{a=0}.
\end{equation}

In the case of a rotating source defined by the metric (\ref{ellip_curve}) and its junction with Kerr empty space time continuity of the first fundamental form is again satisfied by equation (\ref{cond1}) but continuity of the second fundamental form now gives rise to 
\begin{equation} \label{cond3}
	 \sigma(r_{b},\theta)=0, 
\end{equation}
and so is a more general form of (\ref{cond2}). This equation hold for all values of the velocity parameter $a$ as can be seen by the direct calculation of the pressure term $8\pi \sigma=G^{1}_{1}$ which takes the following form:
\begin{eqnarray}
8 \pi \sigma(r,\theta) &=& - \frac{e^{-2\beta}r^4(2r\alpha_{r}-e^{2\beta}+1)}{(r^2+a^2\cos^2\theta)^3} \nonumber \\
& & - \frac{e^{-2\beta}a^2cos^2\theta(-2\alpha_{r}r^3-2r^2+r^2e^{-2\alpha}+r^2e^{2(\alpha+\beta)})}{(r^2+a^2\cos^2\theta)^3}. \nonumber \\
\label{eq11}
\end{eqnarray}
However note in general that:
\begin{equation}
\varpi(r_{b},\theta) \ne 0, \hspace{0.3in}  p(r_{b},\theta) \ne 0, \hspace{0.3in} \varrho(r_{b},\theta) \ne 0, \hspace{0.3in} u^3(r_{b},\theta) \ne 0.
\end{equation}

%SCHWARZSCHILD INTERIOR SECTION
\section{Schwarzschild Interior solution as a source for Kerr empty space-time}\label{interior}
The above procedure may conveniently be used to demonstrate how the Schwarzschild interior solution expressed in curvature coordinates (\ref{curve}) can be embedded into the metric (\ref{ellip_curve}) for a rotating system which matches the Kerr metric smoothly in accordance with the Darmois junction conditions. The Schwarzschild interior solution, with constant energy density is given by:
\begin{equation}\label{alp_sch}
e^{2\alpha}=\frac{1}{4} \left\{ 3\sqrt{1-\frac{2m}{r_{b}}}-\sqrt{1-\frac{2mr^2}{r^3_{b}}} \right\} ^2,
\end{equation}
\begin{equation}\label{bet_sch}
		e^{-2\beta}=1-\frac{2mr^2}{r^3_{b}},
\end{equation}
where $r=r_{b}$ defines the boundary with the Schwarzschild vacuum solution and the solution is written in a form which satisfies the Darmois junction conditions.
This solution can now be embedded into (\ref{ellip_curve}) coupled with (\ref{fbar}) to (\ref{nbar})   to describe a rotating system. 

It follows from the results of the previous section that the values of  $U^a, \chi^a, \xi^a, \sigma, \varpi, p, \varrho$ can be calculated at the boundary $r=r_{b}$. The results are presented in closed form and as a power series for which $a^2<<r^2$ and it is found that: 
\begin{equation}\label{sh1}
		\sigma(r_{b},\theta)=0, \hspace{0.5in} \varpi(r_{b},\theta)=0.
\end{equation}
Moreover:
\begin{eqnarray}
8\pi p(r_{b}, \theta) &=& -\frac{3m\sqrt{r_{b}^2+a^2-2r_{b}m}\sqrt{r_{b}^2-2r_{b}m+a^2\cos^2\theta}}{(r_{b}-2m)(r_{b}^2
+a^2\cos^2\theta)^2} \nonumber \\
  & & + \frac{3m(r_{b}^2+a^2-2r_{b}m)}{(r_{b}-2m)(r_{b}^2+a^2\cos^\theta)^2} \nonumber \\
  &=&- \frac{3ma^2\sin^2\theta}{2r_{b}^4(r_{b}-2m)} + \cdots .\label{pb}
\end{eqnarray}
This confirms that the boundary pressure becomes important only when terms in $a^2$ are significant. The boundary value of the internal density is:
\begin{eqnarray}
8\pi \varrho(r_{b}, \theta) &=& \frac{3m\sqrt{r_{b}^2+a^2-2r_{b}m}\sqrt{r_{b}^2-2r_{b}m+a^2\cos^2\theta}}{(r_{b}-2m)(r_{b}^2
+a^2\cos^2\theta)^2} \nonumber \\
  & & + \frac{3m(r_{b}^2+a^2-2r_{b}m)}{(r_{b}-2m)(r_{b}^2+a^2\cos^\theta)^2} \nonumber \\
  &=&\frac{6m}{r^3_{b}}+\frac{3a^2m((16m-7r_{b})\cos^2\theta+3r_{b})}{2r_{b}^5(r_{b}-2m)}+ \cdots ,\label{rhob}
\end{eqnarray}
and the constant density term for the non-rotating system is apparent when a power series expansions is employed. Furthermore from (\ref{dens}):
\begin{equation} \label{ratio}
\left \{ \frac{U^3}{U^4} \right \}_{r=r_{b}}=\frac{a\sqrt{r_{b}^2-2r_{b}m+a^2\cos^2\theta}}{2(r_{b}^2+a^2\cos^2\theta)(r_{b}^2+a^2-2r_{b}m)},
\end{equation}
which means that:
\begin{eqnarray}
U^{4}(r_{b},\theta) &=& \left(1-\frac{2m}{r_{b}}\right)^{-1/2}-\frac{a^2(8m-r_{b}\sin^2\theta)}{8r_{b}^{3/2}(r_{b}-2m)^{3/2}}+\cdots, \nonumber \\
U^{3}(r_{b},\theta) &=& \frac{a}{2r_{b}^{3/2}\sqrt{r_{b}-2m}}+\cdots. \label{u34}
\end{eqnarray}
The junction conditions are completed by calculating the anisotropy vectors $\chi^a$ and $\xi^a$ at $r=r_{b}$ as follows:
\begin{equation}\label{aneqn}
\chi^a(r_{b},\theta)=\sqrt{ \frac{r^2_{b}-2r_{b}m+a^2}{r^2_{b}+a^2\cos^2\theta}} \delta^a_{1}. \hspace{0.3in}\xi^a(r_{b},\theta)=\sqrt{\frac{1}{r^2_{b}+a^2\cos^2\theta}}\delta^a_{2}.
\end{equation}

%EQUATION OF STATE SECTION
\section{Rotation with equation of state}\label{rstate}
Although in the previous section closed form calculations were given at the junction $r=r_{b}$ no attempt was made to write the closed expressions for the physical parameters in terms of $r$. In general the expressions are too cumbersome and series approximations or numerical calculations would normally be necessary. 

Even for the particular solution due to Tolman \cite{tol} satisfying the equation of state $p=\varrho/3$ where:
\begin{equation}\label{parts} 
		e^{2\alpha}=cr, \hspace{0.5in} e^{-2\beta}=\frac{4}{7},
\end{equation}
the corresponding rotating solution, not matching empty space-time, is most easily expressed as a power series in angular velocity parameter $a$ that provided $a<<r$.  It is straight  forward to show that:
\begin{eqnarray}
	8\pi \sigma&=&\frac{1}{7r^2}+a^2\cos^2\theta 
	\left(-\frac{4}{7cr^5} +\frac{9}{7r^4} 
	- \frac{c}{r^3}   \right ) +\cdots, \nonumber \\
	8\pi \varpi&=&\frac{1}{7r^2}+a^2\cos^2\theta 
	\left(\frac{4}{7cr^5} -\frac{11}{7r^4} 
	+ \frac{c}{r^3}   \right ) +\cdots ,\nonumber  \\	\label{press}
\end{eqnarray}
and interesting to note that the presence of rotation (together with large values of $r$) tends generally to decrease pressure $\sigma$ in the direction of $\chi^a$ whilst at the same time increasing pressure $\varpi$ in the direction of $\xi^a$. In addition the pressure $p$ is 
\begin{equation}\label{pret}
		8\pi p = \frac{1}{7r^2} +a^2\left(-\frac{9}{7cr^5} +\frac{3}{r^4} -\frac{7c}{4r^3} \right ) 
	 +a^2\cos^2\theta \left(\frac{13}{7cr^5} -\frac{32}{7r^4} 
	+\frac{11c}{4r^3}   \right ) + \cdots,   
\end{equation}
with a non-cosinusoidal term which reduces $p$ in the direction orthogonal to  $\chi^a$ and $\xi^a$ in the presence of rotation. The internal density $\varrho$ is:
\begin{equation}\label{det}
	8\pi \varrho = \frac{3}{7r^2} +a^2\left(\frac{3}{7cr^5} +\frac{1}{r^4} -\frac{7c}{4r^3} \right )
	 +a^2\cos^2\theta \left(\frac{5}{7cr^5} -\frac{32}{7r^4} 
	+\frac{19c}{4r^3}   \right ) + \cdots , 
\end{equation}
which generally increases in the presence of rotation.
The expressions for the velocity four vector $U^a$ become:
\begin{equation}\label{ut}
	U^4 = \frac{1}{\sqrt{cr}} +\frac{a^2}{\sqrt{cr}}\left(-\frac{7c}{4r} +\frac{13}{4r^2} -\frac{3}{2cr^3} \right ) +\frac{a^2}{\sqrt{cr}}\cos^2\theta \left(\frac{7c}{4r} -\frac{11}{4r^2} 
	+\frac{1}{cr^3}   \right ) +\cdots,	
\end{equation}
with:
\begin{equation} \label{u33}
	U^3 = \frac{a}{2r^2}\left(\frac{3}{\sqrt{cr}} -\frac{7\sqrt{cr}}{2} \right ) + \cdots.
\end{equation}
Finally the anisotropy vectors $\chi^a$ and $\xi^a$ are given by:
\begin{equation}\label{ctol}
		\chi^a=\left( \frac{4}{7\sqrt{cr}}+\frac{2a^2}{7r^3\sqrt{cr}}-\frac{2a^2cos^2\theta}{7r^2\sqrt{cr}} \right)\delta^a_{1}\cdots, \hspace{0.2in}\xi^a=\left( \frac{1}{r}-\frac{a^2cos^2\theta}{2r^3}\right)\delta^a_{2}+\cdots.
\end{equation}
Clearly the anisotopy deceases with increasing values of $r$.

%Discussion
\section{Discussion}
The research presented here has built upon the notions of axial ellipsoidal symmetry and anisotropy of pressure to present a method of determining rotating interior fluid solutions which match the Kerr (or Kerr-de Sitter) vacuum solution smoothly. These interiors may be thought of as standard base static solutions of Einstein's equations  but now endowed with rotation. The base solutions can be either isotropic or anisotropic but presented in a curvature coordinate system. These base solutions,  also known as R-Solutions can be transformed into T-solutions, applicable within the Schwarzschild radius and it has been shown how these can also be endowed with rotation. The method was applied to empty space-time to give the T-region counterpart of the Kerr and the Kerr-de Sitter solution.

For this method to be applicable it has been necessary to extend the ideas of anisotropy to a three component perfect fluid as metrics with axial ellipsoidal symmetry are not compatible with perfect fluid sources or the two component perfect fluid originally derived by Lettelier. Thus if this technique is to have wide applicability it is necessary to establish the existence of this type of anisotropy from a physical point of view and also the bounds on the validity of the assumption of axial ellipsoidal symmetry. Clearly if rotation is to be based upon a perfect fluid source then space-times  with ellipsoidal symmetry are not appropriate and alternative methods should be employed.

In principle the rotating interiors presented here can be expressed in closed form and an example of this has been given using the Schwarzschild interior metric endowed with rotation at the fluid boundary with empty space-time. None the less the closed forms have a highly complex structure and the use of power series in the angular velocity parameter seem inevitable. Clearly it is important to establish criteria by which to analyse these rotating solutions with a view to establishing models that are physically realistic.

\end{document}